\documentclass[journal]{IEEEtran}
\ifCLASSINFOpdf
\else
\fi
\usepackage{multirow}
\usepackage{flushend}
\usepackage{graphicx}
\usepackage{color}
\usepackage{epstopdf}
\usepackage[square,comma,sort&compress,numbers]{natbib}
\usepackage{amsmath}
\usepackage{amssymb}
\usepackage{algpseudocode,algorithm}
\graphicspath{{Figures/}}

\DeclareMathSizes{10}{8}{7}{6}
\IEEEaftertitletext{\vspace{-2\baselineskip}}
\begin{document}
\title{Multiple Access Technologies for cellular M2M Communications: An Overview}
\author{Mahyar~Shirvanimoghaddam and Sarah~J.~Johnson

\thanks{The authors are with the School of Electrical Engineering and Computer Science, The University of Newcastle, NSW, Australia (e-mail: \{mahyar.shirvanimoghaddam, sarah.johnson\}@newcastle.edu.au).}}
\maketitle

\begin{abstract}
This paper reviews the multiple access techniques for machine-to-machine (M2M) communications in future wireless cellular networks. M2M communications aims at providing the communication infrastructure for the emerging Internet of Things (IoT), which will revolutionize the way we interact with our surrounding physical environment. We provide an overview of the multiple access strategies and explain their limitations when used for M2M communications. We show the throughput efficiency of different multiple access techniques when used in coordinated and uncoordinated scenarios. Non-orthogonal multiple access is also shown to support a larger number of devices compared to orthogonal multiple access techniques, especially in uncoordinated scenarios. We also detail the issues and challenges of different multiple access techniques to be used for M2M applications in cellular networks.

\end{abstract}
\IEEEpeerreviewmaketitle

\section{Introduction}
Machine-to-machine (M2M) communications is expected to become an integral part of cellular networks in the near future. In M2M communications a large number of multi-role devices, such as sensors and actuators, wish to communicate with each other and with the underlying data transport infrastructure. To enable such a massive communication in wireless networks major shifts from current protocols and designs are necessary \cite{MTCSurvey}. That is current wireless networks which have been mainly designed and engineered for human-based applications, such as voice, video, and data, cannot be used for M2M communications due to the different nature of their traffic and service requirements \cite{M2MMInternet}. These differences have posed many questions and challenges in the communication society, in both industry and research sectors.

M2M communications aims at providing the communication infrastructure for emerging Internet of Things and involve the enabling of seamless information exchange between autonomous devices without any human intervention. M2M devices can be either stationary, such as smart meters, or mobile, such as fleet management device, and they can connect to the network infrastructure using either wired or wireless links. Key challenges of massive M2M communications can be listed as follows \cite{M2M_Ericsson2}:

\textbf{\emph{Device cost}}- for the mass deployment of M2M communications low cost devices are necessary for most use cases.

\textbf{\emph{Battery life}}- most M2M devices are battery operated and replacing batteries is not practical for many applications.

\textbf{\emph{Coverage}}- deep indoor and regional connectivity is a requirement for many applications.

\textbf{\emph{Scalability}}- network capacity must be easily scaled to handle a large number of devices forecasted to arise in the near future.

\textbf{\emph{Diversity}}- cellular systems must be able to support diverse service requirements for different use cases, ranging from static sensor networks to tracking systems.

The wired solutions include cable, xDSL, and optical fiber, and can provide high reliability, high data rate, short delay, and high security. However, they are cost ineffective and do not support mobility and scalability; therefore, not appropriate for M2M applications \cite{M2M_Ericsson2}. On the other hand, Wireless capillary (i.e., short range) solutions, such as WLAN and ZigBee, can provide low cost infrastructure and scalability for most M2M applications, but they suffer from small coverage, low data rate, weak security, and severe interference. Wireless cellular, i.e., GSM, GPRS, 3G, LTE-A, WiMAX, etc., however offers excellent coverage, mobility and scalability support, and good security, and the fact that the infrastructure already exists makes it a promising solution for M2M communications \cite{M2M_Ericsson2}. Therefore, our focus in this paper is on wireless cellular solutions for M2M communications.

The mobile industry is standardizing several low power technologies, such as extended coverage GSM (EC-GSM), LTE for machine-type communication (LTE-M), and narrow band IoT (NB-IoT). Since GSM is still the dominant mobile technology in many markets, it is expected to play a key role in the IoT due to its global coverage and cost advantages. EC-GSM enables coverage improvements of up to 20 dB with respect to GPSRS on the 900MHz band \cite{3GPPLTEA}. It is achieved by defining new control and data channels mapped over legacy GSM which provides combined capacity of up to 50000 devices per cell on a single transceiver. LTE-M brings new power saving functionality suitable for serving a variety of IoT applications, which extend battery life to 10 years or more. NB-IoT is a self contained carrier that can be deployed with a system bandwidth of 200 kHz. These initiatives were undertaken in 3GPP Release 13 for M2M specific applications \cite{M2M_Ericsson2}.

Despite all these efforts, further improvements is required in the way that devices communicate with the base station to support a large number of devices and not jeopardizing the human-based communication quality. The multiple access (MA) techniques has been identified as a key area where improvements for M2M communications are needed. The fact that the radio access strategy in LTE is still based on random access mechanisms turns it into a potential bottleneck for the performance of cellular networks when the number of M2M devices grows \cite{IsRandomLTE}. Moreover, radio resources are orthogonally allocated to the users/devices in the current LTE standards, which is not effective for M2M communications when the number of devices goes very large, due to the limited number of radio resources \cite{CoverageM2MLTE}.

In this paper, we consider several multiple access technologies and show their performance in coordinated and uncoordinated scenarios. Overall, coordinated strategies outperform uncoordinated ones as in  coordinated strategies the base station can optimally allocate the radio resources between the devices and support a larger number of devices. We also show that the non-orthogonal multiple access (NOMA) scheme, achieves the highest throughput in both coordinated and uncoordinated strategies, whereas frequency division multiple access (FDMA) has comparable performance in coordinated scenarios. This suggests that FDMA can be effectively used in coordinated scenarios to achieve maximum throughput (this has been considered by 3GPP for M2M communications in the NB-IoT solution), while in uncoordinated scenarios NOMA strategies must be considered to effectively support a large number of devices and use the available radio resources in an efficient manner.

The remainder of the paper is organized as follows. Section II represents the uniques characteristics of M2M communications and its challenges in cellular networks. In Section III, we provide an overview on different multiple access technologies. Coordinated and uncoordinated MA techniques are represented in Section IV and V, respectively, where we characterize their maximum achievable throughput. Practical issues for implementing MA techniques for M2M communications are presented in Section VI. Finally, Section VII concludes the paper.

\section{M2M Communications: Characteristics and Challenges}
Until recently, cellular systems have been designed and engineered for human based applications, such as voice, video, and data, with a higher demand on downlink. M2M communications however have different traffic characteristics, which includes small and infrequent data generated from a very large number of devices, which impose a higher traffic volume on the uplink. In addition, M2M applications have very diverse service requirements. For instance, in alarm signal applications, a small-size message must be delivered to the base station (BS) within 10 msec, while in other applications, such as smart metering, the delay of up to several hours or even a day is tolerable \cite{IoTSmartCity}.

Due to limited radio resources and the large number of devices involved in M2M communications, wireless networks should minimize the time wasted due to collisions or exchanging control messages. The throughput must be large enough to support a large number of devices. Control overhead must be minimized as the payload data in many M2M applications is of small size and the control overhead of conventional approaches in current cellular systems results in an inefficient M2M communications \cite{HybridRAandDataM2M}. In fact, if the control overhead of a protocol is large, the effective throughput is degraded even though the physical data rate may not be affected. It is also required that the effective throughput remain high irrespective of the traffic level \cite{ASurveyofMAC}.

Scalability is another challenge in M2M communications as it is expected that a large number of devices arise in M2M scenarios. These devices have dynamic behaviour, i.e., enter and leave the network frequently; thus the network must easily tolerate the changes in the node density with little control information exchange. Energy efficiency is also one of the most important challenges in M2M communications, as devices in many M2M applications are battery operated and long life times are expected for these devices \cite{M2MRA}. More specifically, the energy spent on radio access and data transmission in M2M communications must be minimized to improve the energy efficiency in a large scale. For instance, in high load scenarios, exchanging control information may consume more than 50\% of the total energy in IEEE 802.11 MAC protocol, which shows its ineffectiveness in dense M2M applications \cite{ASurveyofMAC}.

In many M2M applications, the network latency is a critical factor that determines the effectiveness of the service. For instance in intelligent transportation systems and healthcare monitoring, it is highly important to make the communication reliable and fast. Channel access delay then needs to be minimized to reduce the overall latency in M2M communications. Moreover, in cellular systems, human-to-human (H2H) devices coexist with M2M devices, and the communication protocol must be designed in such a way to not jeopardize the quality of human-based communications. Resource management and allocation are challenging tasks in M2M communications which coexist with H2H applications, as H2H applications have completely different service requirements \cite{ChallM2MAccess}.

These unique characteristics of M2M communications introduce a number of networking challenges in cellular networks. The fundamental issue arises from the fact that most M2M applications involve a huge number of devices. The question is then how the available radio resources have to be shared among devices such that their service requirements are simultaneously met.

\section{Multiple Access Techniques for M2M Communications: A General Overview}
Multiple access techniques can be divided into two broad categories, depending on how the radio resources are allocated to the devices. These include, i) \emph{uncoordinated}, where the devices transmit data using slotted random access and there is no need to establish dedicated resources, and ii) \emph{coordinated}, where devices transmit on separate resources pre-allocated by the base station. In coordinated MA, the base station knows a priori the set of devices that have data to transmit. The BS can also acquire channel state information (CSI) of these devices based on which it allocates resources to optimize system throughput. CSI to the devices can be obtained by each device sending an upload pilot signal.

\begin{figure}[t]
  \centering
  \includegraphics[scale=0.75]{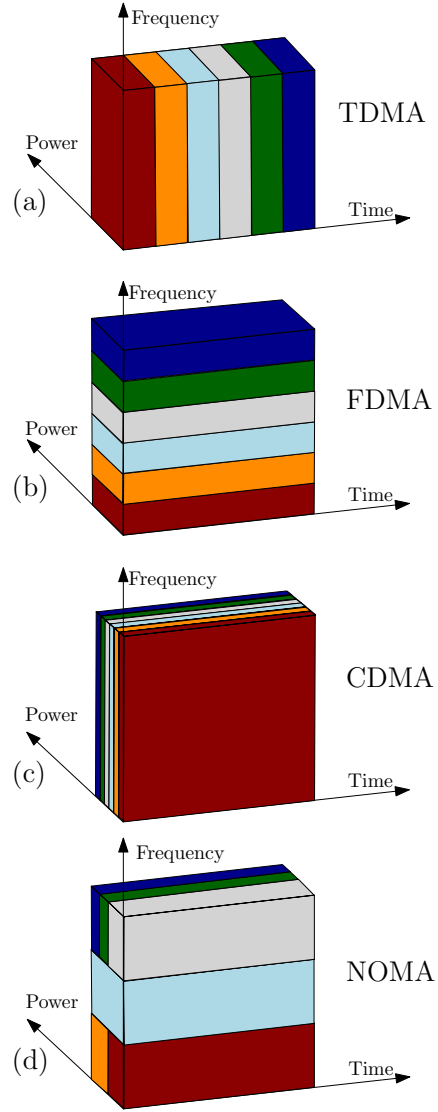}
  \caption{Different multiple access schemes}
  \label{fig:MAschemes}
\end{figure}

Multiple access techniques can be also divided into orthogonal and Non-orthogonal approaches. In orthogonal MA (OMA), radio resources are orthogonally divided between devices, where the signals from different devices are not overlapped with each other. Instances of OMA (see Fig. \ref{fig:MAschemes}) are time division multiple access (TDMA), frequency division multiple access (FDMA), orthogonal frequency division multiple access (OFDMA), and single carrier FDMA (SC-FDMA). First and second generation cellular systems are mainly developed using OMA approaches, which avoid intra-cell interference and simplify air interface design. However, OMA approaches have no ability to combat the inter-cell interference; therefore careful cell planning and interference management techniques are required to solve the interference problem \cite{China_NOMA}.

Non-orthogonal MA (NOMA) techniques have been adopted in second and third generation cellular systems. NOMA allows overlapping among the signals from different devices by exploiting power domain, code domain, and interleaver pattern. Code division multiple access (CDMA) is the well-known example of NOMA which has been adopted in second and third generation cellular systems. CDMA is robust against inter-cell interference, but suffers from intra-cell interference \cite{China_NOMA}. CDMA is also not suitable for data services which require high single-user rates. Rather than CDMA which exploits code domain, NOMA in general exploits power domain. NOMA is also shown to provide better performance than OMA \cite{China_NOMA}. In NOMA, signals from multiple users are superimposed in the power-domain and successive interference cancellation (SIC) is used at the BS to decode the messages. It is also shown that NOMA can achieve the multiuser capacity region both in the uplink and downlink \cite{China_NOMA}.

In this paper, we compare NOMA and OMA strategies in both coordinated and uncoordinated scenarios, and show that NOMA can provide the system with higher capacity to support M2M devices, especially in the uncoordinated scenario. This is achieved by exploiting the power domain, rather than frequency-domain or time-domain as in FDMA and TDMA, respectively.

For the analysis in this paper, we consider a single cell centered by base station and devices uniformly distributed around it in a circular region with radius $R$. The uplink load seen by the base station is modeled by a Poisson point process with mean $\lambda$ arrivals per second. We further assume a time slotted system with a slot duration of $\tau_s$. We perform our analysis on a typical radio resource with slot duration $\tau_s$ and bandwidth $W$. Each device packet is assumed to have a payload of $L$ bits.

The channel from a device located at distance $r$ from the base station is modelled by $g=(r/R)^{-\gamma}$, where $\gamma$ denote the path loss exponent and we ignore shadowing and small scale fading \cite{PowerEfficient}. The received signal-to-noise ratio (SNR) for a device transmitting with power $P_t$ over bandwidth $W_{t}$ is then given by \cite{FunThroM2M}:
\begin{align}
\mu_r=\frac{P_t}{P_{\max}}\frac{W}{W_t}\mu g,
\end{align}
where $P_{\max}$ is the maximum transmit power and $\mu$ is the reference SNR, defined as the average received SNR from a device transmitting at maximum power $P_{\max}$ over bandwidth $W$ located at the cell edge. Without loss of generality, we assume ordered channel gain $g_1\ge g_2\ge \cdots\ge g_K$, where $K$ is the number of devices.

\section{Coordinated Multiple Access Strategies}
In this section, we consider the coordinated multiple access strategies, i.e., TDMA, FDMA, and NOMA, and compare their throughput efficiency. In this section, we assume that the BS has perfect CSI to all the devices.
\subsection{Optimal Throughput FDMA Strategy}
In FDMA, the spectrum is partitioned between the devices and each device will transmit in a portion of the spectrum. Fig. \ref{fig:MAschemes}-b shows the FDMA strategy, where the whole spectrum has been divided between 6 devices, and each device will use its allocated bandwidth for the data transmission.

Using Shannon's capacity formula, the minimum bandwidth required for the transmission of $L$ bits by the $i^{th}$ device over time $\tau_s$ is given by the solution of the following equation \cite{PowerEfficient}:
\begin{align}
\frac{L}{\tau_s W_{\min_i}}=\log_2\left(1+\mu\frac{W}{W_{\min_i}}g_i\right).
\end{align}
The maximum load that can be supported in a resource block of duration $\tau_s$ and bandwidth $W$ is given by:
\begin{align}
K_{\max}=\max\left\{K: \sum_{i=1}^{K}{W_{\min_i}}\le W\right\}.
\end{align}

\subsection{Optimal Throughput TDMA Strategy}
In TDMA, the whole spectrum is used by each device in separate time instances. Fig. \ref{fig:MAschemes}-a shows the TDMA scheme, where the same time duration is allocated for 6 devices, and each device will only transmit in its allocated time slot using the whole spectrum. TDMA is an interesting MA strategy due to its simplicity, but it is not efficient for M2M applications with a large number of devices. Moreover, with increasing the number of devices, each device's transmission will be delayed which is not appropriate for delay-sensitive M2M applications.

Assuming a capacity approaching code and using Shannon's capacity equation, the time required for a device located at distance $r$ from the base station to deliver its packet to the destination is given by \cite{PowerEfficient}:
\begin{align}
\tau\ge \frac{L}{W\log_2(1+\mu_r)},
\end{align}
and the minimum time required to deliver the message is obtained when the device is transmitting with full power $P_{\max}$:
\begin{align}
\tau_{\min_i}=\frac{L}{W\log_2(1+\mu g_i)}.
\end{align}
Similar to FDMA, the maximum number of devices which can be supported in a resource block of duration $\tau_s$ and bandwidth $W$ can then be found as follows:
\begin{align}
K_{\max}=\max\left\{K : \sum_{i=1}^{K}\tau_{\min_i}\le \tau_s\right\}.
\end{align}

\subsection{Optimal Throughput NOMA Strategy}
Unlike TDMA and FDMA, devices in the NOMA strategies are assumed to transmit in the same resource block and their transmissions interfere with each other. We assume that the BS perform successive interference cancellation (SIC), where it starts the decoding with the device with the largest channel gain and treats the signals from other devices as additive noise. After decoding the first device, its signal will be removed from the received signal and the BS continues the decoding for the second device and treats the remainder as additive noise. This process is continued until all the devices are successfully decoded. Under this decoding strategy, the Shannon Capacity formula for the $i^{th}$ device is given by:
\begin{align}
L=W\tau_s\log_2\left(1+\frac{P_i\mu g_i}{1+\sum_{j=i+1}^{K}P_j\mu g_j}\right),
\end{align}
and the required transmit power can be calculated as follows:
\begin{align}
P_i\mu g_i=\left(2^{\frac{L}{W\tau_s}}-1\right)\left(1+\sum_{j=i+1}^{K}P_j\mu g_j\right).
\end{align}

By substituting, $i=K$, we have:
\begin{align}
P_K=\frac{2^{\frac{L}{W\tau_s}}-1}{\mu g_K},
\end{align}
and by going backwards and finding the transmit power for the $i^{th}$ device, we have:
\begin{align}
P_i=\frac{2^{\frac{(K-i)L}{W\tau_s}}\left(2^{\frac{L}{W\tau_s}}-1\right)}{\mu g_i}.
\end{align}

The maximum load that the BS can support in a resource block of bandwidth $W$ and duration $\tau_s$, can be found as follows:
\begin{align}
K_{\max}=\max\left\{K: P_i\le P_{\max}~\text{for}~ i=1,2,...,K\right\}.
\end{align}
\subsection{Comparison between Coordinated MA Techniques}
\begin{figure}[t]
\centering
\includegraphics[width=\columnwidth]{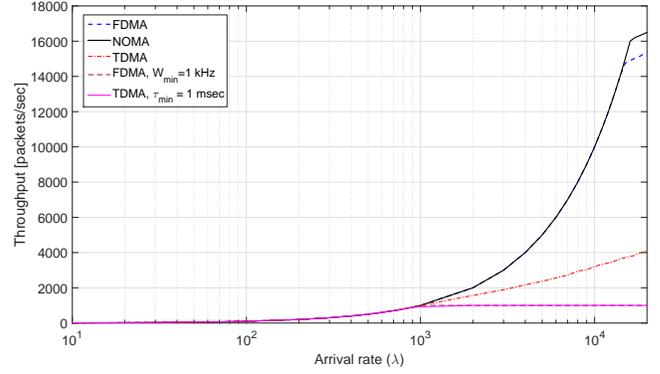}
\caption{Average throughput versus the arrival rate for different coordinated MA techniques. Total available bandwidth is $W=1$ MHz, time slot duration is $\tau_s=1$ sec, and the packet length is $L=1000$ bits. }
\label{fig:MAcomp}
\end{figure}

Fig. \ref{fig:MAcomp} shows the maximum throughput versus arrival rate for different coordinated MA techniques. As can be seen in this figure, NOMA can achieve very high throughput when the arrival rate is very large. FDMA performs very close to the NOMA strategy and can support all the active device for the arrival rates up to 14000 packets per second. The advantage of NOMA comes from the fact that the devices can use the whole spectrum thus achieving a higher throughput compared to FDMA, where only a fraction of the spectrum is used by each device. Also, TDMA cannot support many devices which shows that it not an effective MA strategy for M2M communications.

It is clear that the time slot duration $\tau_i$ and subchannel bandwdith $W_i$ cannot be arbitrarily small in TDMA and FDMA, respectively. As can be seen in Fig. \ref{fig:MAcomp}, if we put some constraints on the minimum time slot duration or subchannel bandwidth, the number of devices which can be supported by FDMA and TDMA would be limited. For example, if the minimum time slot duration for TDMA is set to be $1$ msec, the maximum number of devices which can be supported in a time slot of duration 1 sec is 1000. Similarly, if the minimum suchannel bandwidth in FDMA is set to be 1 kHz, the maximum number of devices which can be supported by the BS will be 1000. This shows that in practical systems where the minimum subhannel bandwidth and time slot duration cannot be very small, the maximum throughput of TDMA and FDMA will be limited. In such cases, NOMA can bring more benefits to the system as it can support a larger number of devices without dividing the radio resource into subchannels or time slots.
\section{Uncoordinated Multiple Access Strategies}
In this section, we assume that the base station does not have CSI to the devices, which is particularly the case for M2M communications with a large number of devices, where it is almost impractical for the base station to estimate the channel to every device with random activities. The only information we assume is available at the BS has, is the traffic load which can be obtained using different load estimation algorithms. 

\subsection{Uncoordinated FDMA}
In this scheme, we assume that the base station chooses a selection probability $p_c$ and broadcasts this information to the devices. Each device which has data to transmit only switches on its transmitter with probability $p_c$. We refer to these devices as active devices. Let $N_c$ denote the number of active device. We further assume that the BS uniformly divides the spectrum into $N_w$ sub-channels, and each device randomly choose a sub-channel for its transmission. We also assume that each device only transmits on a selected sub-channel if the maximum transmit power required to deliver its message to the BS is less than $P_{\max}$, assuming no collision on the selected sub-channel. More specifically, the $i^{th}$ device is transmitting in a sub-channel if the following condition holds:
\begin{align}
\left(2^{\frac{LN_w}{W\tau_s}}-1\right)\le N_w\mu g_i,
\end{align}
therefore, the probability that a device is transmitting can be calculated as follows:
\begin{align}
p\left(\left(2^{\frac{LN_w}{W\tau_s}}-1\right)\le N_w\mu g_i\right)=\left(\frac{N_w\mu}{2^{\frac{LN_w}{W\tau_s}}-1}\right)^{\frac{2}{\gamma}},
\end{align}
which is due to the fact that the devices are uniformly distributed in the cell and the probability that a device is located at distance $r$ is given by $\frac{2r}{R^2}$. The average number of active devices which can deliver their messages, considering no collision, can be found as follows:
\begin{align}
N_p=N_c\left(\frac{N_w\mu}{2^{\frac{LN_w}{W\tau_s}}-1}\right)^{\frac{2}{\gamma}}
\end{align}

As the devices randomly choose a sub-channel for their transmission, more than one device can select the same sub-channel which leads to collision, and the base station cannot decode any of the devices which are simultaneously transmitting on that particular sub-channel. The probability of collision can be calculated as follows \cite{FunThroM2M}:
\begin{align}
P_c=1-\left(1-\frac{1}{N_w}\right)^{N_p-1},
\label{colprob}
\end{align}
and the average number of devices which can successfully deliver their messages to the BS is given by $N_P (1-P_c)$. We assume that the BS finds the optimal values for $p_c$ and $N_w$ such that the number of devices which can be supported by the BS is maximized.

\subsection{Uncoordinated TDMA}
Similar to FDMA, we assume that the BS assigns an access probability $p_c$ to the devices. Let $N_c$ denote the number of active device. We further assume that the BS uniformly divides the time into $N_t$ time slots, and each device randomly chooses a time slot for its transmission. We also assume that the each device only transmits in a selected time slot if the maximum transmit power required to deliver its message to the BS is less than $P_{\max}$, assuming no collision on the selected time slot. More specifically, the $i^{th}$ device is transmitting in a time slot, if the following condition holds:
\begin{align}
\left(2^{\frac{LN_t}{W\tau_s}}-1\right)\le \mu g_i,
\end{align}
therefore, the probability that a device is transmitting can be calculated as follows:
\begin{align}
p\left(\left(2^{\frac{LN_t}{W\tau_s}}-1\right)\le \mu g_i\right)=\left(\frac{\mu}{2^{\frac{LN_t}{W\tau_s}}-1}\right)^{\frac{2}{\gamma}},
\end{align}
which is due to the fact that the devices are uniformly distributed in the cell and the probability that a device is located at distance $r$ is given by $\frac{2r}{R^2}$. The average number of active devices which can deliver their messages, considering no collision, can be found as follows:
\begin{align}
N_p=N_c\left(\frac{\mu}{2^{\frac{LN_t}{W\tau_s}}-1}\right)^{\frac{2}{\gamma}}.
\end{align}

The average number of devices which can successfully deliver their messages to the BS is given by $N_P (1-P_c)$, where $P_c$ is given by (\ref{colprob}) by replacing $N_w$ with $N_t$. We assume that the BS finds the optimal values for $p_c$ and $N_t$ such that the number of devices which can be supported by the BS is maximized.
\subsection{Uncoordinated NOMA}
We consider that each device performs power control such that the received SNR at the BS for each device is $\gamma_0$. A device will only transmit if and only if the transmit power required to achieve the SNR $\gamma_0$ at the base station is less than $P_{\max}$. Let $N_p$ denote the number of devices which can transmit, i.e., their required transmit power is less than $P_{\max}$. The achievable rate for the devices considering the successive interference cancellation at the BS can be calculated as follows:
\begin{align}
R_{\min}=\log_2\left(1+\frac{\gamma_0}{1+(N_p-1)\gamma_0}\right),
\label{rmin}
\end{align}
and a message of length $L$ can be delivered by $N_p$ devices if $W\tau_s R_{\min}\ge L$. Using (\ref{rmin}), the required SNR $\gamma_0$ to successfully deliver a message of length $L$ at the BS is derived as follows:
\begin{align}
\gamma_0=\frac{1}{\frac{1}{2^{\frac{L}{W\tau_s}}-1}-N_p},
\end{align}
and accordingly the number of devices which can be supported at the BS is upper bounded as follows:
\begin{align}
N_{p}\le\frac{1}{2^{\frac{L}{W\tau_s}}-1}
\end{align}
\begin{figure}[t]
\centering
\includegraphics[width=\columnwidth]{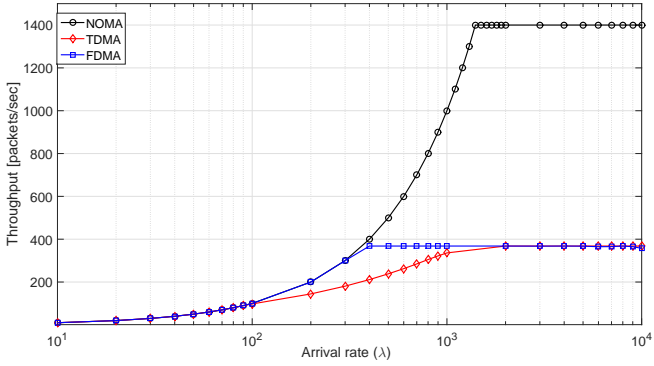}
\caption{Average throughput versus the arrival rate for different uncoordinated MA techniques. Total available bandwidth is $W=1$ MHz, time slot duration is $\tau_s=1$ sec, and the packet length is $L=1000$ bits. The minimum time slot duration for TDMA is considered to be 1 ms and the minimum subchannel bandwidth in FDMA is considered to be 1 kHz. }
\label{fig:UncComp}
\end{figure}
\subsection{Comparison between uncoordinated MA techniques}
Fig. \ref{fig:UncComp} shows the maximum number of devices which can be supported by the base station versus different arrival rates for uncoordinated MA strategies. The minimum time slot duration for TDMA is considered to be 1 ms, which corresponds to $N_t=1000$, and the minimum subchannel bandwidth in FDMA is considered to be 1 kHz, which corresponds to $N_w=1000$. As can be seen in this figure, NOMA can support much larger number of devices compared to the FDMA and TDMA strategies. This is due to the high collision probability in uncoordinated FDMA and TDMA in high arrival rates, while in NOMA a large number of devices can simultaneously transmit at the same resource block by exploiting the power domain. This shows the advantage of NOMA in uncoordinated scenarios which can be an excellent choice for M2M applications with a large number of devices and random traffic. Moreover as can be seen in Fig. \ref{fig:UncComp}, FDMA outperforms TDMA in moderate loads but they perform similarly in low and high arrival rates.

It is important to note that in NOMA the constraint on minimum time slot duration or subchannel bandwidth do not affect the throughput efficiency. This is due to the fact that in NOMA all the devices are transmitting in the whole bandwidth in all slot duration. One could consider some limitations in the minimum power difference between the devices, which mostly depends on the hardware capability to distinguish different power levels which is out of scope of this paper. 
\section{Practical Considerations of Massive NOMA for M2M Communications and Future Directions}
NOMA can bring many benefits to cellular systems which include, but are not limited to, the following. NOMA can effectively use the spectrum and provide higher throughput by exploiting power domain and non-orthogonal multiplexing. It also provides robust performance gain in high mobility scenarios. NOMA is also compatible with OFDMA and can be applied on top of OFDMA for downlink and SC-FDMA for uplink. It can be also combined with multi-antenna techniques to improve the system performance. Using NOMA, multiple users can simultaneously transmit in the same subband without being identified by the destination a priori. The devices can attach their terminal identities to their messages and the base station can identify the devices after decoding their messages. The RA procedure can be eliminated and therefore the access delay and signaling overhead will be significantly reduced \cite{China_NOMA}.

Although NOMA can improve spectrum efficiency and system capacity, there are many practical challenges for this technology to be potentially  used in real wireless systems for M2M communications. Here, we outline the main practical consideration of massive NOMA for M2M communications.

First, in uncoordinated strategies the base station needs to estimate the arrival rate to effectively detect the devices. In uncoordinated FDMA, the BS needs to know the number of devices to find the optimal access probability and the number of subbands. In NOMA, the problem is much more complicated as the BS runs the SIC and needs to know the number of devices with different power levels. For simplicity, one could consider that the devices perform power control such that only one power level is received at the BS, but this may have some implications on the actual performance of the system as the overall system data rate will be dominated by the device with the lowest SINR; and thus will not effectively use the available spectrum. However, even with this simplification and suboptimal power allocations, NOMA outperforms FDMA in uncoordinated scenarios and can support a large number of devices under high loads.


Second, channel estimation at the devices in necessary in uncoordinated strategies employing NOMA techniques. This is due to the fact that the devices are not identified by the BS beforehand and they are simultaneously transmitting at the same resource block. To enable the BS to detect the devices and decode their messages, the devices need to perform channel estimation and adjust their power so the BS only deals with some known power levels rather than unknown channel gains. On the other hand, to effectively perform SIC, the multipath effect must be carefully taken into consideration as multipath will spread the signal over time, which decreases the effective signal to noise ratio for each device, and makes the BS unable to perform SIC. One can consider several techniques, such as time reversion \cite{GreenTRDMA}, to eliminate the multipath effect by treating the channel between each device and the BS as the natural match filter. This has been shown an effective way to combat multpath effect for several fixed location M2M applications \cite{IoT_TRDMA}.

Third, NOMA requires synchronization among the devices at the symbol level. This is very challenging as providing time synchronicity between a large number of devices distributed in a large environment is tedious. However, the devices in many M2M applications are deployed in fixed locations, so each device can determine its propagation delay using different distance estimation strategies or using control information periodically sent by the BS.

Fourth, as the number of devices transmitting in each resource block in uncoordinated NOMA is random, the physical data rate cannot be determined beforehand. One could consider a very low rate code at each device, but it might be inefficient when used in low-to-moderate loads. An effective strategy is then to use rateless codes to automatically adapt to the traffic condition. Authors in \cite{Mahyar_TWC} have proposed to use analog fountain codes to enable massive multiple access for M2M communications and achieve very high throughput even in high loads. Moreover, as shown in \cite{Mahyar_SPM_Raptor}, binary rateless codes can be effectively used to enable NOMA for M2M communications. These coding strategies were mainly proposed to maximize the throughput in M2M communications and for delay sensitive applications with very short messages more advanced coding techniques should be combined with rateless ideas to enable low latency massive multiple access in M2M communications.

Last but not least, NOMA is still in its early stage of its development and more research work must be done to clearly identify its effectiveness in real scenarios. From an information theoretic point of view, it achieves the capacity region of the multiple access channel and thus is optimal in terms of throughput. But in real M2M applications when NOMA is jointly considered with medium access control layer in real world scenarios, it might not be as efficient as OMA techniques, which have been considered as effective multiple access techniques for a long time and several issues and challenges have been solved over the years.

\section{Conclusions}
In this paper, we provided an overview of multiple access techniques for emerging machine-to-machine communications in cellular systems. The unique challenges of M2M communications were represented, where we identified scalability, energy efficiency, and reliability, as the most important features for every potential multiple access technology which is considered for M2M communications. We provided a simple study on the throughput efficiency of multiple access techniques in both coordinated and uncoordinated scenarios. NOMA was shown to provide the highest throughput in both coordinated and uncoordinated scenarios, whereas FDMA has shown comparable performance with NOMA in coordinated scenarios. NOMA is shown to be scalable in uncoordinated scenarios and can support a large number of devices. It can be also  combined with different access management schemes to control the load over the base station. We also provided some of the practical issues in NOMA which needed to be considered for the use of NOMA strategies for M2M communications in future cellular systems.

\bibliographystyle{IEEEtran}
\footnotesize
\bibliography{IEEEabrv,dp-16}

\begin{thebibliography}{10}
\providecommand{\url}[1]{#1}
\csname url@samestyle\endcsname
\providecommand{\newblock}{\relax}
\providecommand{\bibinfo}[2]{#2}
\providecommand{\BIBentrySTDinterwordspacing}{\spaceskip=0pt\relax}
\providecommand{\BIBentryALTinterwordstretchfactor}{4}
\providecommand{\BIBentryALTinterwordspacing}{\spaceskip=\fontdimen2\font plus
\BIBentryALTinterwordstretchfactor\fontdimen3\font minus
  \fontdimen4\font\relax}
\providecommand{\BIBforeignlanguage}[2]{{%
\expandafter\ifx\csname l@#1\endcsname\relax
\typeout{** WARNING: IEEEtran.bst: No hyphenation pattern has been}%
\typeout{** loaded for the language `#1'. Using the pattern for}%
\typeout{** the default language instead.}%
\else
\language=\csname l@#1\endcsname
\fi
#2}}
\providecommand{\BIBdecl}{\relax}
\BIBdecl

\bibitem{MTCSurvey}
H.~Shariatmadari, R.~Ratasuk, S.~Iraji, A.~Laya, T.~Taleb, R.~Jantti, and
  A.~Ghosh, ``Machine-type communications: Current status and future
  perspectives toward {5G} systems,'' \emph{{IEEE} Commun. Mag.}, vol.~53,
  no.~9, pp. 10--17, Sep. 2015.

\bibitem{M2MMInternet}
G.~Wu, S.~Talwar, K.~Johnsson, N.~Himayat, and K.~Johnson, ``{M2M}: From mobile
  to embedded internet,'' \emph{{IEEE} Commun. Mag.}, vol.~49, no.~4, pp.
  36--43, 2011.

\bibitem{M2M_Ericsson2}
\BIBentryALTinterwordspacing
``{Cellular networks for massive IoT},'' Ericsson, Tech. Rep. Uen 284 23-3278,
  January 2016. [Online]. Available:
  \url{https://www.ericsson.com/res/docs/whitepapers/wp\_iot.pdf}
\BIBentrySTDinterwordspacing

\bibitem{3GPPLTEA}
``{T}hird {G}eneration {P}artnership program ({3GPP}): {S}ervice {R}equirements
  for {M}achine-{T}ype {C}ommunications ({MTC}); {S}tage 1,'' Technical
  Specification 22.368, {V}.13.0.0, Jun. 2014.

\bibitem{IsRandomLTE}
A.~Laya, L.~Alonso, and J.~Alonso-Zarate, ``Is the random access channel of
  {LTE} and {LTE-A} suitable for {M2M} communications? a survey of
  alternatives,'' \emph{{IEEE} Commun. Surveys Tuts.}, vol.~16, no.~1, pp.
  4--16, First 2014.

\bibitem{CoverageM2MLTE}
G.~Naddafzadeh-Shirazi, L.~Lampe, G.~Vos, and S.~Bennett, ``Coverage
  enhancement techniques for machine-to-machine communications over {LTE},''
  \emph{{IEEE} Commun. Mag.}, vol.~53, no.~7, pp. 192--200, Jul. 2015.

\bibitem{IoTSmartCity}
A.~Zanella, N.~Bui, A.~Castellani, L.~Vangelista, and M.~Zorzi, ``Internet of
  things for smart cities,'' \emph{IEEE Internet of Things Jounral}, vol.~1,
  no.~1, pp. 22--32, Feb. 2014.

\bibitem{HybridRAandDataM2M}
D.~Wiriaatmadja and K.~W. Choi, ``Hybrid random access and data transmission
  protocol for machine-to-machine communications in cellular networks,''
  \emph{{IEEE} Trans. Wireless Commun.}, vol.~14, no.~1, pp. 33--46, Jan. 2015.

\bibitem{ASurveyofMAC}
A.~Rajandekar and B.~Sikdar, ``A survey of {MAC} layer issues and protocols for
  machine-to-machine communications,'' \emph{IEEE Internet of Things Journal},
  vol.~2, no.~2, pp. 175--186, April 2015.

\bibitem{M2MRA}
M.~Hasan, E.~Hossain, and D.~Niyato, ``Random access for machine-to-machine
  communication in {LTE}-advanced networks: issues and approaches,''
  \emph{{IEEE} Commun. Mag.}, vol.~51, no.~6, pp. 86--93, 2013.

\bibitem{ChallM2MAccess}
K.~Zheng, S.~Ou, J.~Alonso-Zarate, M.~Dohler, F.~Liu, and H.~Zhu, ``Challenges
  of massive access in highly dense {LTE}-advanced networks with
  machine-to-machine communications,'' \emph{{IEEE} Wireless Commun. Mag.},
  vol.~21, no.~3, pp. 12--18, Jun. 2014.

\bibitem{China_NOMA}
A.~Li, Y.~Lan, X.~Chen, and H.~Jiang, ``Non-orthogonal multiple access ({NOMA})
  for future downlink radio access of {5G},'' \emph{China Communications},
  vol.~12, no. Supplement, pp. 28--37, Dec. 2015.

\bibitem{PowerEfficient}
H.~S. Dhillon, H.~C. Huang, H.~Viswanathan, and R.~A. Valenzuela,
  ``Power-efficient system design for cellular-based machine-to-machine
  communications,'' \emph{{IEEE} Trans. Wireless Commun.}, vol.~12, no.~11, pp.
  5740--5753, Nov. 2013.

\bibitem{FunThroM2M}
------, ``Fundamentals of throughput maximization with random arrivals for
  {M2M} communications,'' \emph{{IEEE} Trans. Commun.}, vol.~62, no.~11, pp.
  4094--4109, Nov. 2014.

\bibitem{GreenTRDMA}
B.~Wang, Y.~Wu, F.~Han, Y.~H. Yang, and K.~J.~R. Liu, ``Green wireless
  communications: A time-reversal paradigm,'' \emph{{IEEE} J. Sel. Areas
  Commun.}, vol.~29, no.~8, pp. 1698--1710, Sep. 2011.

\bibitem{IoT_TRDMA}
Y.~Chen, F.~Han, Y.~H. Yang, H.~Ma, Y.~Han, C.~Jiang, H.~Q. Lai, D.~Claffey,
  Z.~Safar, and K.~J.~R. Liu, ``Time-reversal wireless paradigm for green
  internet of things: An overview,'' \emph{IEEE Internet of Things Jounral},
  vol.~1, no.~1, pp. 81--98, Feb. 2014.

\bibitem{Mahyar_TWC}
M.~Shirvanimoghaddam, Y.~Li, M.~Dohler, B.~Vucetic, and S.~Feng,
  ``Probabilistic rateless multiple access for machine-to-machine
  communication,'' \emph{{IEEE} Trans. Wireless Commun.}, vol.~14, no.~12, pp.
  6815--6826, Dec. 2015.

\bibitem{Mahyar_SPM_Raptor}
\BIBentryALTinterwordspacing
M.~Shirvanimoghaddam, S.~J. Johnson, and M.~Dohler, ``An efficient massive
  access strategy based on superposition {R}aptor codes for {M2M}
  communications,'' \emph{CoRR}, 2016. [Online]. Available:
  \url{http://arxiv.org/pdf/1602.05671v1.pdf}
\BIBentrySTDinterwordspacing

\end{thebibliography}

\begin{IEEEbiographynophoto}{Mahyar Shirvanimoghaddam}
received the B. Sc. degree with 1’st Class Honours from University of Tehran, Iran, in September 2008, the M. Sc. Degree with 1’st Class Honours from Sharif University of Technology, Iran, in October 2010, and the Ph.D. degree from The University of Sydney, Australia, in January 2015, all in Electrical Engineering. He then held a research assistant position at the Centre of Excellence in Telecommunications, School of Electrical and Information Engineering, The University of Sydney, before coming to the University of Newcastle, Australia, where he is now a Postdoctoral Research Associate at the School of Electrical Engineering and Computer Science. His general research interests include channel coding techniques, cooperative communications, compressed sensing, machine-to-machine communications, and wireless sensor networks.
\end{IEEEbiographynophoto}
\begin{IEEEbiographynophoto}{Sarah Johnson}
received the B.E. (Hons) degree in electrical engineering in 2000, and PhD in 2004, both from the University of Newcastle, Australia. She then held a postdoctoral position with the Wireless Signal Processing Program, National ICT Australia before returning to the University of Newcastle where she is now an Australian Research Council Future Fellow. Sarah’s research interests are in the fields of error correction coding and network information theory. She is the author of a book on iterative error correction published by Cambridge University Press.
\end{IEEEbiographynophoto}
\end{document}